# Processor Based Active Queue Management for providing QoS in Multimedia Application

N. Saravana Selvam. M.E.,
Professor/Department of CSE,
Sree Sowdambika College of Engineering, Aruppukottai,
Tamil Nadu – 626 134, India.
saravanaselvam.kcet@gmail.com,

Dr. S. Radhakrishnan,
Senior Professor & Head/ Department of CSE,
Arulmigu Kalasalingam College of Engineering,
Krishnankoil, Tamil Nadu - 626 190, India.
srk@kalasalingam.ac.in

*Abstract—The objective of this paper is to implement the Active Network based Active Queue Management Technique for providing Quality of Service (QoS) using Network Processor(NP) based router to enhance multimedia applications. The performance is evaluated using Intel IXP2400 NP Simulator. The results demonstrate that, Active Network based Active Queue Management has better performance than RED algorithm in case of congestion and is well suited to achieve high speed packet classification to support multimedia applications with minimum delay and Queue loss. Using simulation, we show that the proposed system can provide assurance for prioritized flows with improved network utilization where bandwidth is shared among the flows according to the levels of priority. We first analyze the feasibility and optimality of the load distribution schemes and then present separate solutions for non-delay-sensitive streams and delay-sensitive streams. Rigorous simulations and experiments have been carried out to evaluate the performance.*

## 1. INTRODUCTION

QoS in the existing and emerging applications in the Internet has been a big challenge to the Internet Programmer. Real time applications such as audio/video conversations and on demand movies are interactive, have delay constraints or require high throughput. There were many mechanisms proposed in the past to enable the internet to support applications with varying service requirements. These include admission control, queue management algorithms and scheduling. Admission control is meant to prevent network overload so as to ensure service for applications already admitted. It is deployed at the network edges where a new flow is either admitted, when network resources are available or rejected otherwise. Active queue management has been proposed to support end-to-end congestion control in the Internet. The aim is to anticipate and signal congestion before it actually occurs.

Scheduling determines the order in which the packets from different flows are served. The above mechanisms are building blocks of Quality of Service (QoS) architectures.The end-to-end congestion control in the Internet requires some form of feedback information from the congested link to the source of data traffic, so that they can adjust their rates of sending the data. Therefore explicit feedback mechanism, such as Active Queue Management algorithms can be used to control congestion in the network [1,2].

Multimedia traffic has become high due to the growth of multimedia applications such as video streaming, video conferencing, IP telephony and e-learning. These Multimedia applications require better Quality of service which can be provided by properly classifying and scheduling the packets, shaping the Traffic, managing the bandwidth, and reserving the resources. All these techniques require the fundamental task Packet classification at the router, to categorize the packets into flows.

A QoS-enabled network is composed of various functions for providing different types of service to different packets such as rate controller, classifier, scheduler, and admission control. The scheduler function of them determines the order in which packets are processed at a node and/or transmitted over a link. The order in which the packets are to be processed is determined by the congestion avoidance and packet drop policy (also called *Active Queue Management*) at the node. Here Dynamic QoS (DQoS) can be applied regardless of existing network settings.

## 2. ACTIVE NETWORKS

In an Active network, the router or switches of the network perform customized computation on the messages flowing through them. For example, the user of an active network could see a "trace" program to each router arranged for the program to be execute when their packets are processed.

Each Active packet, called Capsule, carries both the program fragment and data parts. The program part is stored within packet options that are recognized only on those active nodes and triggers the interruption to be performed on the nodes. Normal IP nodes ignore the program part and treat these capsules as normal IP packets.

A flow is a sequence of packets and can be uniquely identified by the information placed in packet headers, which are source and destination IP addresses, source and destination port number and protocol. Active packets will not interfere with routing protocols. Alternative routes are possible as long as defined in the local forwarding table. Active packets are used for QoS provisioning in conjunction with and as a complement to existing frameworks in order to optimize the usage of existing QoS capabilities.

A. Active Network Approaches

A number of different approaches have been proposed to implement the active networks architecture.

i. There will be some standard services or modules in the network node, which will be selected and invoked through options carried in the user's packets. The rest of the user's packets will be treated as data to be processed by the invoked routine. The current IP processing modules can be thought of as an example of such an approach.

ii. The "programmable switch" model: this approach allows user code to execute on the network nodes. The code is downloaded into the nodes out-of-band, and the normal flow packets are treated as data are input to this node. Depending upon the node's security policy, such code downloads can be restricted only to system administrators.

iii. The "Capsule" model: Each packet is treated as a full-fledges program, to be executed at each intermediate nodes. This is the most general of the three approaches discussed.

## 3. ACTIVE QUEUE MANAGEMENT

The implementation of AQM for emergency purposes seeks to drop the packets of non-emergency traffic as much as needed to allow as much emergency traffic as possible to proceed through the router. The goal is to provide lower packet loss for emergency traffic so that emergency communications can proceed more dependably. It should be noted, however, that emergency traffic in general does not need to have better delay or jitter performance than normal traffic. Emergency traffic is no different than non-emergency traffic in that sense. The use of AQM, therefore, is meant to create lower packet loss for emergency traffic by acting as a filtering mechanism to perform prioritized packet discarding. The filtering also seeks, however, to provide the best service possible to non-emergency traffic by not unnecessarily favoring emergency traffic.

A. Queue Management

Queues represent locations where packets may be held (or dropped). Packet scheduling refers to the decision process used to choose which packets should be serviced or dropped. Buffer management refers to any particular discipline used to regulate the occupancy of a particular queue. At present, support is included for drop-tail (FIFO) queuing, RED buffer management, CBQ (including a priority and round-robin scheduler), and variants of Fair Queuing including, Fair Queuing (FQ), Stochastic Fair Queuing (SFQ), and Deficit Round-Robin (DRR). In the common case where a *delay* element is downstream from a queue, the queue may be *blocked* until it is re-enabled by its downstream neighbor. This is the mechanism by which transmission delay is simulated. In addition, queues may be forcibly blocked or unblocked at arbitrary times by their neighbors (which is used to implement multi-queue aggregate queues with inter-queue flow control).

B. Need for Active Queue Management

The technique for managing router queue lengths is to set maximum length (n terms of packets) for each queue. It accepts packet for the queue until the maximum length is reached, then rejects or drops the incoming packets until the queue (length) decreases, as a packet from the queue has been transmitted. This technique is known as drop tail (FIFO).It has the following drawbacks:

1. Lock-Out :
In some situations tail drop allows a single connection. i.e. it does not allow any other connections.

2. Full Queue :
Once the queue is full, the packets are dropped. So it is important to reduce steady state queue size and this is one of the important goals of queue management.

Active Queue management mechanism can provide following advantages for responsive flows:

a. Reduce the number of packets dropped in routers in packet switched networks; the arrival of packets at destination is bursty in nature. If all the queue space in a router is already committed to "steady state" traffic, then the router cannot handle the buffer bursts. Hence Active queue management maintains small average queue size, which in turn facilitates minimizing the dropping of packets.

b. Provide low delay interactive service. By specifying small average queue size, queue management will reduce the delays seen in the packet flows in a network. This effect can be observed in interactive applications such as video on demand [1].

C. Queuing Components for QoS

This section will give an overview over traffic conditioning components in general. Quality of Service is a kind of service discrimination. A network component (e.g. a router) has to handle packets in different manners in order to achieve a certain Quality of Service for a specific flow or type of packets. This differentiation between packets may be done on a per flow basis as in RSVP or for an aggregation of packets, distinguished by different DSCP (Differentiated Services Code Point) values as it is done by Differentiated Services. Within the Internet the processing speed of network components is assumed high compared to the bandwidth of the links between them. Congestion occurs when a router has to transmit more packets over a link than the link's capacity allows. The router has to discard packets. All approaches to realize QoS are based on dropping the"right" packets. To handle packets differently, an Internet router usually has a queuing system attached to his outgoing interfaces. Simple queuing systems like FIFO (first in first out) queues are capable to intercept short bursts of packets exceeding the output bandwidth. More complicated queuing systems allow handling packets differently, putting them to different kind of queues and processing these queues with different priority. Several mechanisms were developed for various purposes.

## 4. INTEL IXP2400 NETWORK PROCESSOR

The advent of network processors was driven by an increasing demand for high throughput and flexibility in packet routers. As a first step in the evolution from software-based routers to network processors, the bus connecting network interface cards with the central control processor (Control Plane, CP) was replaced by a switch fabric. As demand grew for even greater bandwidth, network interface cards were replaced by Application-Specific Integrated Circuits (ASICs), meaning that packets no longer had to be sent to the CP for forwarding.

### a. Background

The Intel® IXP2400 is designed to perform a wide range of functionalities such as multi-service switches, routers, broadband access devices and wireless infrastructure systems. It implements a high-performance parallel processing architecture on a single chip that is suitable for processing complex algorithms, detailed packet inspection, traffic management and forwarding at the wire speed. The architecture of an integrated network processor IXP 2400 NP shown in Figure 1.

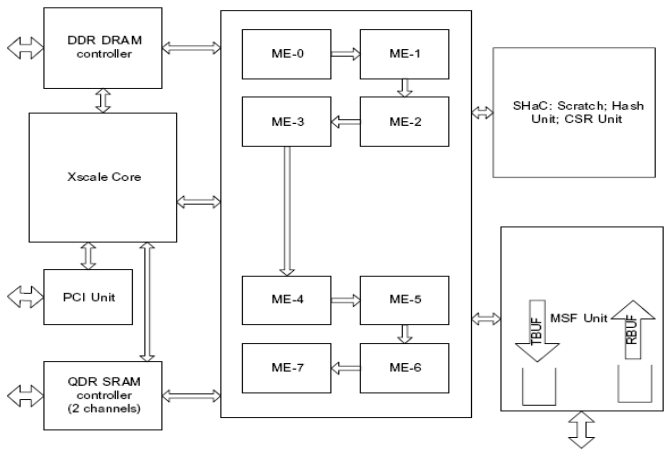

Figure: 1 The Block Diagram of an IXP 2400

It has a single 32-bit XScale core processor, eight 32-bit MicroEngines (MEs) organized as two clusters, standard memory interfaces and high speed bus interfaces. Each microengine has 256 general purpose registers that are equally shared between eight threads. Microengines exchange information through an on-chip scratchpad memory or via 128 special purpose next neighbor registers. Data transferring across the MEs and locations external to the ME, (for eg DRAMs, SRAMs etc.) are done by the available 512 Transfer Registers.

The Xscale core is responsible for initializing and managing the chip, handling control and management functions and loading the ME instructions. Each ME has eight hardware-assisted threads of execution with no context switch overhead. All threads in a particular ME execute code from the instruction stored on that ME, whose size is 4K. The SRAMs and DRAM are off chips that are shared by all processors. In general, SRAM is used for storing the table information such as routing table and DRAM is used for packet buffering. Also the IXP2400 chip has a pair of buffers (BUF), Receive BUF (RBUF) and Transmit BUF (TBUF) that are used to send / receive packets to / from the network ports with each of size 8 Kbytes. The data in RBUF and TBUF is divided into sub blocks referred to as elements.

## 5. IMPLEMENTATION

Packets of non-real time application like the packets for e-mail, file transfer, etc cannot tolerate loss but can afford delay By keeping all these issues in mind priority and scheduling is maintained by programming a IXP2400, to work with buffering, located in the physical memory or DDR-RAM of Intel Network Processor .

The Workbench provides packet simulation of media bus devices as well as simulation of network traffic. There is a feature under the Packet Simulation called Packet Simulation Status that enables the observation of the number of packets being received into the media interface and number of packets being transmitted from the fabric switch interface. To make sure that flows are not synchronized, each flow begins its transmission at a random time within the first 40 msec. The QoS setting in D-QoS is automatically adjusted to offer the interruption according to the buffer capacity of the flows.

Media streaming for real-time playback is highly delay-sensitive, and the playback quality is the most important concern to the clients. Too much jitter may greatly deteriorate the playback quality. To reduce the jitter, media units of each stream must be processed timely and orderly in the NPs.

When voice and video packets need to pass through router (IXP 2400) will be treated and processed at the highest priority as it is real time traffic. The first priority will be for the expedited forwarding which handles real time application packets. This minimize the packet delay for the real-time data packets in IPv6 DiffServ Network to maximize the throughput of the data like voice, video etc in which delay is not tolerable. When EF traffic got some break then second priority for assured forwarding traffic will be forwarded.

In this process both incoming traffic and stored packets in the memory will be moved if there is any. The third priority packets are buffered accordingly, when there are no packets or traffic for EF or AF then the packets for third priority means best effort will be forwarded. The framework for the implementation shown in the Figure 2.

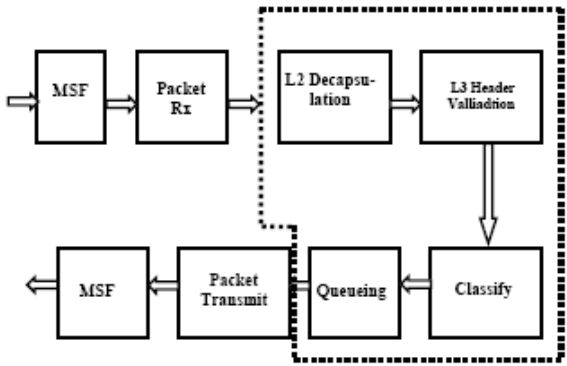

Figure 2:  Framework for the implementation

The simulator is configured to generate and inject packet streams to ports 0 & 1 of the device.(Figure 3) Packet Receive Microengine has been interfaced with the Media Switch Fabric Interface (MSF). The packets are injected through Media switch Fabric Interface and Receive Buffer (RBUF) reassembles incoming mpackets (packets with length 64 bytes). For each packet, the packet data is written to DRAM, the packet meta-data (offset, size) is written to SRAM. The Receive Process is executed by microengine (ME0) using all the eight threads available in that microengine (ME0). The packet sequencing is maintained by executing the threads in strict order (Figure 4). The packet buffer information is written to a scratch ring for use by the packet processing stage Communication between pipeline stages is implemented through controlled access to shared ring buffers. (Figure 5)

[Figure: 3 Packet Simulation statuses table]

Figure: 3 Packet Simulation statuses

The forwarding functionality can be implemented in two ways, through *fast* path and through *slow* path. In fast path, a hashing table is maintained for known data streams to enable fast packet processing. The hashing table is updated at fixed interval of 50ms so that the dynamic performance of the algorithm is maintained. While in slow path, calculations are performed for per-packet basis to find the best possible next-hop at that instant.

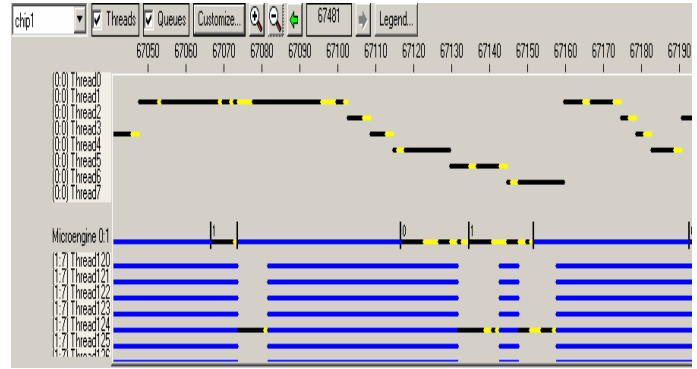

Figure: 4    Execution of Thread in the Microengine

[Figure: 5 Data stored in shared ring buffers table]

Figure: 5   Data stored in shared ring buffers

Classifier microblock is executed by microengine (ME1) that removes the layer-3 header from the packet by updating the offset and size fields in the packet Meta descriptor. The packets are classified into different traffic flows based on the IP header and are enqueued in the respective queue. Then the Packet Transmit microblock segments a packet into m-packets and moves them into TBUFs for transmitting the packets over the media interface through different ports. The egress packet streams are forwarded to the output ports 0-4 of the device.

Multimedia stream is artificially simulated with UDP flows since voice and video applications use UDP rather than TCP. Traffic generated in this paper/work consists of RTP/UDP packets, UDP packets (with large TTL and with less TTL) and TCP packets. All these packets are uniformly distributed in the traffic. Traffic is generated with a constant rate of 1000Mb/s and inter-packet gap is set to 96 ns.

In such an NE, upon receipt of a PDU, the following actions occur. The header of the PDU is parsed (HdrP) and the type of the PDU is identified (PktID) to decide which forwarding table to search (e.g., virtual local area network (VLAN), multiprotocol label switching (MPLS), L2/L3). The PDU is classified (CL) to establish the flow and class of the PDU and possibly its priority. The packet Transmitter micro block monitors the MSF to see if the TBUF threshold for specific ports has been exceeded. If the TBUF value for the output port 0 (VoIP pkts) exceeds, here we are allowing the packets to transmit until it reaches the MAX TBUF value instead of dropping the packets at the tail end. if it still reaches the max threshold (85% of the o/p buffer) starts forwarding it to the other ports which are dedicated to low priority flows.

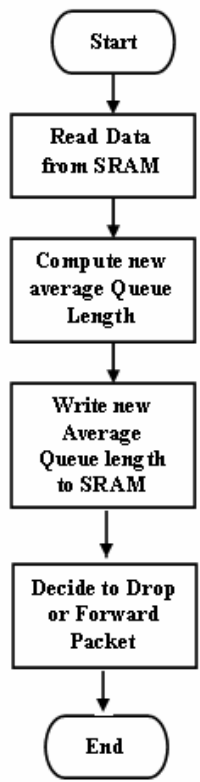

Figure: 6. AQM before Ordered thread execution

If it happens for other non priority flows it stops transmitting on that port and any requests to transmit packets on that port are queued up in local memory and if the data rate increases further the packets are dropped at the end. (Figure 6) The Packet Transmitter micro block periodically updates the classifier with information about how many packets have been transmitted.

If no further treatment is required then the AQM function decides whether to forward or discard the PDU(Protocol Data Unit) according to the output queues_ fill levels and the PDU's class and/or priority and possibly its color. The scheduler (SCH) ensures that the PDU's transmission is prioritized according to its relative class/priority. The traffic shaper (TS) ensures that the PDU is transmitted according to its QoS profile and/or agreements.

The queuing mechanism used within this interruption mode aims to give higher priority to the privileged flow. Other flows previously classified into DiffServ's EF (time-sensitive) and AF(non-time-sensitive) classes are transmitted with the lowest priority. To provide different services for time-sensitive (EF) flows and non-time sensitive (AF) flows, the two lowest priority levels are reserved for each one of these two flow types. The time-sensitive flows previously classified into DiffServ's EF class are placed in the second lowest priority level, while other flows previously belonging to DiffServ's AF classes are given the lowest priority level. Thus, EF traffic receives lower-loss, lower-delay and lower-jitter comparing to AF tr

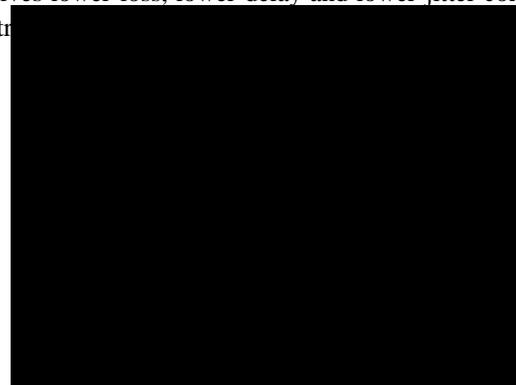

Figure: 7    Packet Loss rate in Queue

Here (Figure 7) the loss rate in the queue decreases drastically when we are applying the proposed queue management technique over the queues present in the egress edge. At the same time packet delay (Figure 8) is also reduces due to the execution of threads in strict order.

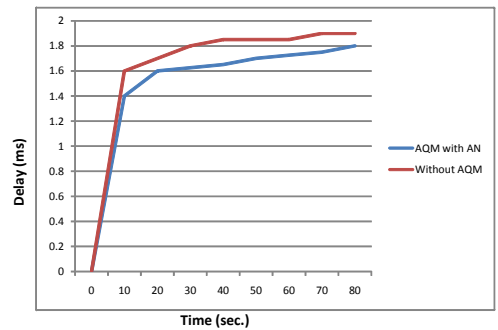

Figure: 8    Packet Queuing Delay

## 6. CONCLUSION

In this paper, we investigated the problem of AQM in multimedia streams on a NP-based active router**.** Some of the open issues in the NP space include software durability, which they share with many other specialized, embedded systems. Therefore, it is currently nearly impossible to write code that would easily port to a different family. For code running in the data plane, use of a smart optimizing compiler permits to write relatively architecture-independent code.

With appropriate, evolving libraries, key sub functions can be transferred progressively and seamlessly from software implementation to hardware, maintaining backwards compatibility. In the control plane, standard interfaces are being developed and joined with capabilities for dynamic NP feature discovery to allow NP-independent code.

Stepping back to see the big picture, we can conclude that versatility is one of NPs strongest features. With comparably little effort, it is possible to implement new features to dramatically increase the value offered by a network device, and to offer new and powerful functions, often combined from amongst a wide variety of existing building blocks. We believe that in the networking world, this makes NPs a strong competitor to ASICs for all but the highest-performance network devices and thus expect their use to grow dramatically in the near future. The simulation results shows that the Active Network technique indeed improves the overall utilization of network resources, and decreases the packet loss rate and queuing delay. Additionally, evaluation results demonstrate that Active Network mechanism is most effective when the high-priority queues suffer serious congestion while resources are still available in low-priority queues.

## REFERENCES

[1] Idris Rai, Ernst W Biersack, "Size Based Scheduling to improve the performance of short TCP Flows", IEEE Network January 2005.

[2] Arnaud Legout and Earnst W, "Revisiting the Fair Queuing Pagadigm for End-to-End Congestion Control", IEEE Network, September 2002.

[3] Andreas Kind, *The Role of Network Processors in Active Networks*, IBM Zurich Research Lab, 2003.

[4] Harish. H. Kenchannavar, U.P. Kulkarni and A.R. Yardi, " A Comparison Study of End-to-End Delay Using Different Active Queue Management Algorithms:, CIMCA 2008.

[5] Jaesung Hong, Changhee Joo and Saewoong Bahk, " Active Queue management algorithm considering queue and load states", Elsevier Computer Network, November 2006.

[6] Enqing Dong and Xiaoqian Ji, " A New Active Queue Management Scheme based on Packet loss Ratio", ICSP 2006.

[7] Bhaskar Reddy. T and Ali Ahammed, " Performance Comparison of Active Queue Management Techniques" Journal of Computer Science, 4(12), 2008.

## AUTHORS PROFILE

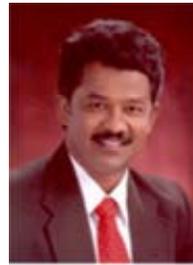

N. Saravanaselvam is working as a Professor in Department of Computer Science and Engineering at Sree Sowdambika College of Engineering, Aruppukottai, Tamilnadu, India. He has completed his B.E. Electronics and Communication Engineering and M.E. Computer Science and Engineering in Arulmigu Kalasalingam College of Engineering, Krishnankoil under Madurai Kamaraj University, Madurai. Now he is a research scholar of Anna University, Chennai. He has guided more than 25 B.E. /M.E. Projects. His field of interest is Network Engineering.

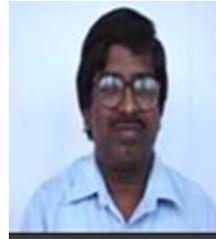

Dr.S.Radhakrishnan is presently working as Senior Professor and Head, Department of Computer Science and Engineering at Arulmigu Kalasalingam College of Engineering, Krishnankoil, TamilNadu, India. He has completed his M.Tech. and Ph.D., in Biomedical Engineering from Institute of Technology, Banaras Hindu University. He has guided more than 50 M.E./M.Tech. Projects and 10 M.Phil. Thesis. Currently ten candidates are working for Ph.D. under his guidance. His fields of interests are Network Engineering, Computer Applications in medicine and Evolutionary computing. He is also serving as Project Director (Network technologies) in TIFAC CORE in Network Engineering at Arulmigu Kalasalingam College of Engineering. He has more than 10 publications to his credit.